\documentstyle{article}

\begin{document}

\title{{\large\bf HIGHER POWER SQUEEZED STATES, JACOBI MATRICES,\\
AND THE HAMBURGER MOMENT PROBLEM}\thanks {Contribution to the {\it 5th International Conference on Squeezed States and Uncertainty Relations} in Balatonf\"{u}red, Hungary, May 27-31, 1997.}}
\author{Bengt Nagel \\
 Division of Theoretical Physics, Royal Institute of Technology \\
 SE-100 44 Stockholm --- Sweden.\\
{\it nagel@theophys.kth.se}} 
\maketitle

\begin{abstract}
{\it k}:th power (amplitude-)squeezed states are defined as the normalized states giving equality in the Schr\"{o}dinger-Robertson uncertainty relation for the pair of the real and the imaginary parts of the {\it k}:th power of the one-mode annihilation operator. Equivalently they are the set of normalized eigenstates (for all possible complex eigenvalues) of the Bogolubov transformed \lq \lq {\it k}:th power annihilation operators\rq \rq  $\: \mu\,a^{k}+\nu\,a^{+k},\:| \mu|^{2}-|\nu|^{2}=1$. Expressed in the number representation the eigenvalue equation leads to a three term recursion relation for the expansion coefficients, which can be explicitly solved in the cases $k=1,2$ . The solutions are  essentially Hermite  and Pollaczek polynomials, respectively. $k=1$ gives the ordinary squeezed states, i.e. displaced squeezed vacua.

For $k\geq3$, where no explicit solution has been found, the recursion relation for the (formal) case $\mu=\nu=1$ defines a Jacobi  matrix related to a classical Hamburger moment problem, which is undetermined. This implies that the symmetric operator $a^{k}+a^{+k}$ has for $k\geq3$ an infinity of self-adjoint extensions, all with disjoint discrete spectra. The corresponding squeezed states are well-defined, however.

\end{abstract}

\section{The Schr\"{o}dinger-Robertson Uncertainty Relation and Corresponding Minimum Uncertainty States}

The {\it Schr\"{o}dinger-Robertson (SR) uncertainty relation} for two (in general) noncommuting hermitian operators $A$ and $B$ in a (normalized) state $\phi$ is
\begin{equation} \label{E1}
\Delta A^{2}\Delta B^{2}\,-\,[\Delta (AB)]^{2} \geq \frac{1}{4}\,|\langle[A,B]\rangle|^{2},
\end{equation}
with the covariance $\Delta(AB)=\langle(A'B'+B'A')/2\rangle,\:A'=A-\langle A\rangle,\:\langle A\rangle=(\phi,A\phi).$\\
This inequality is obtained from Schwarz inequality
\begin{eqnarray*}
(f,f)(g,g)\geq |(f,g)|^{2}=[{\rm Re}(f,g)]^{2}+[{\rm Im}(f,g)]^{2}= \nonumber \\
\{\frac{1}{2} [(f,g)+(g,f)]\}^{2}+\{\frac{1}{2i}[(f,g)-(g,f)]\}^{2},
\end{eqnarray*}
putting  $f=A'\phi,\:g=B'\phi$, and using hermiticity (and some domain conditions, since $A$ and $B$ are in general unbounded operators). Equality holds if and only if $f$ and $g$ are linearly dependent, which means either {\bf 1}. $f$ and/or $g\:=0$, or {\bf 2}. not {\bf1} and $f+cg=0,\:c\neq 0$, complex number. In case {\bf2} we get ${\rm Im}(f,g)={\rm Im}\,c\:(g,g),\:{\rm Re}(f,g)=-{\rm Re}\,c\:(g,g),\:(f,f)=|c|^{2}(g,g)$. If $[A,B]=i\,C,\:C>0$, we get ${\rm Im}(f,g)>0$, i.e. we must be in case {\bf 2}, and ${\rm Im}\,c >0$. $\phi$ cannot be an eigenfunction of $A$ or $B$, but is an eigenfunction of the linear combination $A+cB$. All the quantities in the SR relation (an equality in this case) can then be expressed in terms of $\langle C \rangle $ and $c$. {\rm Re}\,c\,=\,0 corresponds to the covariance $\Delta(AB)=0$, we get a (generalized) Heisenberg uncertainty relation with equality, $\phi$ is a {\it minimum uncertainty state} relative to the operators $(A,B)$. In the general case we could talk about a {\it minimum (SR)-uncertainty state} relative to $(A,B)$. In \cite{trifonov94} it is called a {\it generalized intelligent state}. The set of such states is then the same as the set of all (normalized) eigenfunctions of the operators $A+cB$ for all possible $c$ in the upper complex half plane.

In terms of the non-hermitian combinations ${\sf a}=A+iB,\:{\sf a}^{+}=A-iB$ (\lq \lq generalized annihilation and creation operators\rq \rq) we express the  set of minmum (SR)-uncertainty states as the set of all possible eigenfunctions of the set of \lq \lq Bogolubov transformed generalized annihilation operators\rq\rq \, $\: {\sf a}(\nu)=\muÊ\:{\sf a}+\nu\:{\sf a}^{+},\:\mu=\sqrt{1+|\nu|^{2}},\:\nu$ arbitrary complex. We have $\nu/\mu=(i-c)/(i+c)$, and the commutation relation $[{\sf a}(\nu),{\sf a}(\nu)^{+}]=2C$. The Bogolubov transformation leaves the (operator) value of the commutator invariant.

\section{Amplitude Power Squeezed States}

We now apply the above to the case of ${\sf a}=a^{k}$. With ${\sf a}_{k}(\nu)=\mu\,a^{k}+\nu \, a^{+k}$ we get
 \begin{equation} \label{E2} 
[{\sf a}_{k}(\nu), \, {\sf a}_{k}(\nu)^{+}]=[a^{k},a^{+k}]=(N+k)!/N!-N!/(N-k)!\equiv f_{k}(N)>0. 
\end{equation}
Here $\mu$ and $\nu$ are as before, and $N=a^{+}a$. We are evidently in the situation with a positive operator $C$ described before. Except for the case $k=1$, when $f_{1}(N)=1$, $f_{k}(N)$ is a positive, strictly  increasing function of $N$ (e.g. $f_{2}(N)=4N+2,\: f_{3}(N)= 9N^{2}+11N+6 $). This gives the difference between the case of ordinary squeezed states ($k=1$) and the higher power squeezed states ($k\geq 2$).

In the first case, $k=1$, we get a {\it third equivalent way of defining the set of squeezed states}, as the set of {\it Perelomov coherent states} of the representation of the product group ${\rm H}\times{\rm M}(2,{\rm\bf R})$ of the Heisenberg-Weyl group H and the metaplectic group ${\rm M}(2,{\rm\bf R})$ (the double covering of ${\rm SU}(1,1)$), where one uses the vacuum as isotropy vector. This construction gives as the set of squeezed states the set $\{\, D(\alpha)\,S(\zeta)|0\rangle; \alpha,\zeta=r\cdot e^{i2\vartheta} \in {\rm\bf C}\}$ of displaced squeezed vacuum states; each such state is by construction the vacuum state of a transformed annihilation operator $\mu\, a+\nu \, a^{+} + \beta,\: \mu = \cosh r,\:\nu=\sinh r \cdot e^{i2\vartheta},\:\beta=-(\mu\alpha+\nu\alpha^{*})$ which according to our earlier definition is a minimum (SR)-uncertainty state for the pair $(q,p)$.
The essential theorem used here is the basic result that any two irreducible representations of the canonical commutation relations are unitarily equivalent, so the above transformation to a \lq \lq new\rq\rq \, annihilation operator is unitarily implemented. (See e.g. \cite{nagel95} for more details and references about this construction).

{\it For $k\geq 2$ the Perelomov coherent state construction is not possible}; the linear inhomogeneous transformation leaving the commutation relation invariant cannot be unitarily implemented, since such a unitary operator would have to commute with the right hand side of the commutation relation. This implies that it would have to be a function of $N$ (since all $f_{k}(N), k\geq 2$, have simple spectra), and such a function can only multiply $a^{k}$ with a phase factor.

\section{Solving the Eigenvalue Equation for the k:th Power Squeezed States in the Number Representation}

So we have defined the set of {\it k:th power {\rm (amplitude)} squeezed states} as the set of normalized solutions (up to phase factors) of the eigenvalue equation
\begin{equation} \label{E3}
(\mu \, a^{k}+\nu \, a^{+k})\,|\lambda;\nu\rangle=\lambda\, |\lambda;\nu\rangle; \:\mu=\sqrt{1+|\nu|^{2}},\:\nu \in {\rm \bf C}
\end{equation}
This equation can be treated in (at least) three different representations (reps) of the state vectors: the {\it photon number} (harmonic oscillator excitation number) {\it rep} (the $n$-rep), the ordinary {\it configuration space rep} (the $q$-rep), and the {\it Fock-Bargmann rep} (the $z$-rep). The last rep is closely related to the {\it coherent state rep}. The connection between the reps can be given by the ON bases correspondences
\[
\{|n\rangle\} \leftrightarrow \{u_{n}(q)=N_{n}H_{n}(q) \cdot e^{-q^{2}/2} \} \leftrightarrow \{z^{n}/\sqrt{n!}\},\: N_{n}=(\sqrt{\pi}\,2^{n}n!)^{-1/2}.
\]
The annihilation and creation operators $a,\, a^{+}$ have the forms
\[  a \leftrightarrow (q+d/dq)/\sqrt{2} \leftrightarrow d/dz;\: a^{+} \leftrightarrow (q-d/dq)/\sqrt{2} \leftrightarrow z.  \]
In the $n$- and $z$-reps a suitable scaling transformation reduces the solution of (\ref{E3}) to the formal case $\mu=\nu=1$, i.e. the study of the eigenvalue equation
\begin{equation} \label{E4}
A_{k}|\lambda;k,\kappa \rangle \equiv (a^{k}+a^{+k})|\lambda;k,\kappa\rangle=\lambda|\lambda;k,\kappa\rangle.
\end{equation}
The notation $|\lambda;k,\kappa\rangle$ will be explained below; $\kappa$ takes values $0,..,k-1$.

For $k=1$ Eq. (\ref{E4}) taken in the $n$-rep leads after a simple factor transformation to the well-known recursion relation for Hermite polynomials.

For $k=2$ we have the case of amplitude-squared squeezed states introduced in the first paper in \cite{hillery87}. Some further contributions (not a complete list!) for this case are given in \cite{hillery87}. Since the papers in general use equality in the generalized Heisenberg relation instead of the more general (SR)-relation used here, the results are restricted to the case of real $\nu$. Even with this restriction the general form for the expansion coefficients in the $n$-rep has only been obtained recently \cite{marian97}. Here we obtain, on the basis of results from \cite{nagel95}, the form of these expansion coefficients in terms of (special cases of) Pollaczek polynomials. The relation to the subject in \cite{nagel95} is that in Eq. (\ref{E4}) we have $A_{2}=a^{2}+a^{+2}=4\:J_{1}$, where $J_{1}$ (often called $ K_{1}$) is one of the hyperbolic generators of the standard unitary representation of SU(1,1) in the harmonic oscillator Hilbert space. (In [2] I actually studied another hyperbolic generator, $J_{2}=(a^{2}-a^{+2})/4i $, but this only gives a phase factor difference in each expansion coefficient.) Since the standard representation of SU(1,1) consists of two irreducible representations, actually of the two-fold covering M(2,{\bf R}) of SU(1,1), corresponding to even and odd number states, the spectrum of $A_{2}$ is doubly degenerate. For the self-adjoint operator $A_{2}$ the spectrum is restricted to the real line, with corresponding generalized eigenfunctions, but the amplitude-squared squeezed states are well-defined Hilbert space vectors for any complex $\lambda$ because of the exponentially damping factors $(\nu/\mu)^{m}$.

For $k\geq 3$ the operator $A_{k}$ is in a similar way reduced to each of the subspaces $H_{\kappa}=$ linear span of $\{|m\,k+\kappa\rangle;\,m=0,1,..\}, \:\kappa=0,1,..,k-1$. Whereas for $k=1$ and 2 the solutions of Eqs.Ê(\ref{E3}) and (\ref{E4}) in either representation can be expressed in known transcendental functions,  confluent hypergeometric or hypergeometric  functions, for $k\geq3$ the solutions seem to be more general functions; in the $q$- and $z$-reps we get linear $k$:th order differential equations. In the $n$-rep we get for (\ref{E4}) a three term recursion relation which I have not been able to solve in known polynomials. However, for all $k$ the recursion relation defines an infinite Jacobi matrix (see next Section), which by classical theory is directly related to the Hamburger moment problem (mass distribution on the full real line). It turns out that there is an essential difference between the cases $k=1,2$ and $k\geq3$. In the {\it first two cases the Hamburger problem is determined}, i.e. has a unique solution; this corresponds to the property that the symmetric operator $A_{k}$ (defined  first e.g. on the space of finite linear combinations of number states of each sector $H_{\kappa}$) is {\it essentially self-adjoint}, and its self-adjoint closure has on each $H_{\kappa}$ as simple spectrum the whole real line. For $k\geq 3$ the {\it moment problem  is undetermined}, there are infinitely many solutions to the problem; correspondingly the symmetric operator $A_{k}$, defined as above, has {\it deficiency indices} (1,1) in each subspace. In each subspace we then have a one-parameter family of self-adjoint extensions, each having a purely discrete spectrum on the real line, and different extensions having disjoint spectra. The solutions of Eq. (\ref{E4}) are actually normalizable for any complex $\lambda$.  Considered in the whole Hilbert space  $A_{k}$ has deficiency indices $(k,k)$, which implies that the set of possible self-adjoint extensions in the full Hilbert space can be parametrized by the set of unitary $k\times k$ matrices. The diagonal unitary  matrices are the ones that don't mix the different subspaces in the extension process. --- The non-uniqueness of the extensions of the operator $A_{k}$ of course does not affect the set of $k$:th power squeezed states, which is well-defined, since the recursion relation always has a unique solution for any given (complex) value of $\lambda$.

We use as \lq\lq Ansatz\rq\rq \, for the solution of (\ref{E3}) in  $H_{\kappa}$

\begin{equation} \label{E5}
|\lambda;\nu\rangle=N(\lambda,\nu)\sum_{m=0}^{\infty}(\nu/\mu)^{m/2}f_{m}|mk+\kappa\rangle, \mbox{ to get with}
\end{equation}

\begin{equation} \label{E6}
b_{m,k,\kappa}=[\prod_{p=0}^{k-1}(mk+\kappa+1+p)]^{1/2} \:\approx k^{k/2}m^{k/2}, \, {\rm large}\: m,
\end{equation}
the three term recursion relation (dropping two indices on the $b$:s), and fixing the normalization by taking the first coefficient equal to 1:
\begin{equation} \label{E7}
b_{m}f_{m+1}-\lambda' f_{m}+b_{m-1}f_{m-1}=0\,,\,f_{-1}=0,\,f_{0}=1;\,\lambda'=\lambda/\sqrt{\mu \nu}.
\end{equation}
Observe that evidently the solution to (\ref{E4}) can be written
\begin{equation} \label{E8} 
|\lambda;k,\kappa\rangle=N'(\lambda,k,\kappa) \sum f_{m}|mk+\kappa\rangle,
\end{equation}
where $f_{m}$ solves (\ref{E7}) with $\lambda$ substituted for $\lambda'$.
Since $|\nu/\mu|<1$ the series in (\ref{E5}) will be convergent, and define a vector in Hilbert space, even if (\ref{E8}) is divergent with $f_{m}$ increasing slower than exponentially.

Before analyzing the recursion relation (\ref{E7}) in general, we will solve the case $k=2$.

${\bf k=2}$. $b_{m,2,0}=\sqrt{(2m+1)(2m+2)};\:b_{m,2,1}=\sqrt{(2m+2)(2m+3)}$
Take first $\kappa=0$, even number states. Put
$f_{m}=i^{m}\sqrt{(1/2)_{m}/m!}\, g_{m},\:(a)_{m}=\Gamma(a+m)/\Gamma(a)$, and $\lambda'=4\,\ell$ to get
\begin{equation} \label{E9}
(m+1/2)g_{m+1}+i2\ell g_{m}-mg_{m-1}=0.
\end{equation}
To connect with the treatment in [2] I have introduced $\ell $, which is the eigenvalue of the SU(1,1) generator $J_{1}=(a^{2}+a^{+2})/4$.

Solving (\ref{E9}) with the standard Laplace method for difference equations, or else by comparing with a suitable contiguity relation for the hypergeometric function \cite{erdelyi53}, and using the initial condition $g_{0}=1$ we get $g_{m}=\,_{2}F_{1} (-m,1/4+i\ell,1/2;2)$. As final result we get $f_{m}=P_{m}(\ell,1/4)$, where we have introduced the notation for a special case of the Pollaczek polynomials \cite {erdelyi253} with a new normalization (also compared to the one used in [2])
\begin{equation}
P_{m}(x,b)=i^{m}\sqrt{(2b)_{m}/m!}\:_{2}F_{1}(-m,b+ix,2b;2)
\end{equation}
which for every  $b>0$ are real polynomials of degree {\it m} in {\it x}, even or odd according to the parity of {\it m}, with a positive coefficient of the highest degree term, and with all zeros simple and on the real line. They form a complete orthonormal set of polynomials on the real line with the weight function
\begin{equation}
\rho_{b}(x)=2^{2b-1}|\Gamma(b+ix)|^{2}/\pi\Gamma(2b).
\end{equation}
From the orthonormality and completeness of these polynomials follow, as discussed in [2], the completeness and generalized orthonormality (in that order!) of the set of generalized eigenfunctions of the generator $J_{1}$.

Using the above $f_{m}$ in the definition (\ref{E5}) we can calculate the normalization constant and by various formulas for hypergeometric functions obtain the expressions for the amplitude-squared even number squeezed states also in the $q$- and $z$-reps.

For $\kappa=1$, i.e. odd number states, we obtain in a similar way $f_{m}=P_{m}(\ell,3/4)$.

\section{Jacobi Matrices and the Classical Moment Problem}

We return to the general study of the recursion relation (\ref{E7}). For facts about Jacobi matrices and the classical moment problem we refer to \cite{akhiezer65}.

(\ref{E7}) is the eigenvalue equation $Af=\lambda' f$ for the infinite Jacobi matrix

\begin{equation}
  A= \left (
				\begin{array}{cccc}
					0 & b_{0} & 0 & ...     \\
			    b_{0} & 0 &b_{1} & 0 \\
					 0 & b_{1} & 0 &b_{2}\\
					..  & 0 & b_{2}  & 0     
				\end{array}
				\right)  ,  f=
			\left  (
  \begin{array}{c}
		f_{0}\\
		f_{1}\\
		f_{2}\\
			....
		\end{array}
		\right  ); \mbox{ we have all } b_{m}>0
\end{equation}

Our case is special in the sense that the diagonal matrix elements of $A$ are all zero; for a general Jacobi matrix they can take any (real) values.

The Hamburger moment problem is to find a (positive) measure $\sigma$ on the real line corresponding to the moments $s_{m}=\int_{-\infty}^{\infty}u^{m}d\sigma(u),\: m=0,1,2,..$. A necessary and sufficient condition for the existence of a solution is that the sequence $\{s_{m}\}$ is positive, i.e. that the Hankel forms $\sum_{i,k=0}^{n}s_{i+k}x_{i}x_{k}>0$ for non-zero vectors $ \{x_{0},x_{1},..\}$. This condition can be expressed in the condition that a sequence of determinants of the Hankel forms be positive, and certain combinations of the determinants determine the coefficients $b_{m}$ in a Jacobi matrix; expressed in properties of the associated Jacobi matrix the condition for the solvability of the corresponding moment problem is that all $b_{m}>0$.

The Jacobi matrix belongs to {\it type D} (limit point case; corresponding moment problem is determined, i.e. there is a unique solution for the measure), provided  $\sum1/b_{m}=\infty $ (sufficient but not necessary condition!). Then the symmetric operator $A$, defined e.g. on finite sequences $f$, has a closure which is self-adjoint. It is easily seen that the condition on  $\{b_{m}\}$ is satisfied for $k=1\mbox{ and }2 $.

If the diagonal elements are bounded in absolute value , $b_{m-1}b_{m+1} \leq b_{m}^{2}$  (from some $m_{0}$ on), and $\sum 1/b_{m}<\infty$, then the Jacobi matrix belongs to the other type, {\it type C} (limit circle case), and the above mentioned
closure has deficiency indices (1,1), with properties of the self-adjoint extensions mentioned in Section 3. This holds for $k\geq3$, and the nonuniqueness of the self-adjoint extensions of the operator $A_{k}$ in these cases throws some light on the discussion of the existence of "higher power squeezing generators" in the papers in \cite{fisher84}.  This term refers to the (unfulfilled) expectation  that in the same way as $A_{2}=a^{2}+a^{+2}$  can generate ordinary squeezed vacua by exponentiation in the Perelomov coherent state formulation, the higher order $A_{k}$:s might  generate higher-power squeezed vacua. As we have seen, the Perelomov construction does not work for $k\geq2$. The impossibility of defining  unique self-adjoint forms for  the prospective higher order squeezing generators gives another aspect on this fact.

\end{document}